\documentclass[lineno]{jfm}
\usepackage{graphicx}
\usepackage{newtxtext}
\usepackage{newtxmath}
\usepackage{natbib}
\usepackage{hyperref}
\usepackage{float}
\hypersetup{
    colorlinks = true,
    urlcolor   = blue,
    citecolor  = black,
}

\newcommand{\RomanNumeralCaps}[1]
\linenumbers

\title{Forecasting subcritical cylinder wakes with Fourier Neural Operators}

\author{Peter I Renn\aff{1}
  \corresp{\email{prenn@caltech.edu}},
  Cong Wang\aff{1},
  Sahin Lale \aff{1}, 
  Zongyi Li \aff{1},
  Anima Anandkumar \aff{1},
 \and Morteza Gharib\aff{1}}

\affiliation{\aff{1}Division of Engineering and Applied Science, California Institute of Technology, Pasadena, CA 91125, USA}

\begin{document}

\nolinenumbers
\maketitle

\begin{abstract}
We apply Fourier neural operators (FNOs), a state-of-the-art operator learning technique, to forecast the temporal evolution of experimentally measured velocity fields. FNOs are a recently developed machine learning method capable of approximating solution operators to systems of partial differential equations through data alone. The learned FNO solution operator can be evaluated in milliseconds, potentially enabling faster-than-real-time modeling for predictive flow control in physical systems. Here we use FNOs to predict how physical fluid flows evolve in time, training with particle image velocimetry measurements depicting cylinder wakes in the subcritical vortex shedding regime. We train separate FNOs at Reynolds numbers ranging from $\Rey =$ 240 to $\Rey =$ 3060 and study how increasingly turbulent flow phenomena impact prediction accuracy. We focus here on a short prediction horizon of ten non-dimensionalized time-steps, as would be relevant for problems of predictive flow control. We find that FNOs are capable of accurately predicting the evolution of experimental velocity fields throughout the range of Reynolds numbers tested (L2 norm error $< 0.1$) despite being provided with limited and imperfect flow observations. Given these results, we conclude that this method holds significant potential for real-time predictive flow control of physical systems. 
\end{abstract}


\section{Introduction}
\label{sec:headings}

Vortex shedding in a cylinder wake is among the most fundamental and well-studied problems in fluid mechanics. This phenomenon, otherwise known as a Kármán vortex street, is still relevant for a huge array of industries and applications today. The vortex shedding process has been studied in detail for decades with several comprehensive review papers on the subject, so we will keep our descriptions here brief \citep{williamson1996review}. Vortex shedding is first observed around $\Rey = 50$, at which point it has been observed that an instability occurs in the previously steady recirculation regions of the wake. This instability results in a famously beautiful and well-ordered pattern of laminar alternating vortices convecting downstream away from the cylinder. There has been some experimental variation observed in defining where the laminar vortex shedding regime ends but it is generally placed around $\Rey$ = 190, at which point small-scale three-dimensional instabilities form and the transition to turbulence begins \citep{williamson1996threed}. Following transition, the behavior of cylinder wakes from $\Rey$ = 300 to $\Rey$  = 200 000 was labeled the "irregular range" by \citet{roshko1958bluffbody} (otherwise known as the sub-critical regime). This regime is characterized by increasing three-dimensional effects, irregular velocity fluctuations, and the transition of the outer shear layer ($\Rey$ = 1200) \citep{williamson1996review}. 

Modeling vortical wakes in the ``irregular range'' is possible in standard computational fluid dynamics (CFD) simulations, however, can be computationally expensive and time intensive \citep{pereira2017cylinder}. As an alternative to the conventional numerical solvers widely adopted by the CFD community, there has been growing interest in applications of data-driven methods for modeling the time evolution of fluid flows. In contrast to standard CFD solvers, data-driven methods require data and time up-front but are significantly faster to evaluate once trained. 

Previously proposed data-driven methods for predicting the time evolution of fluid flows primarily focus on neural networks with varying structures (e.g. convolutional layers, deconvolutional layers, long short-term memory cells, and variational autoencoders, etc.). Several of these works involve applying neural networks to predict the evolution of computational simulations of laminar flow around a cylinder \citep{hasegawa2020a,hasegawa2020b,morimoto2021}. \citet{srinivasan2019} and \citet{nakamura2021} apply similar neural network approaches to computational simulations of turbulent shear layers  and three-dimensional channel flows, respectively. In both cases, the data-driven models match statistical quantities of the flow well but struggle to make accurate instantaneous predictions. \citet{han2019} demonstrate a method for predicting the time evolution of cylinder flows from computational simulations at various Reynolds numbers including some turbulent, but require dozens of previous time-steps as input. \citet{wu2022} propose a multi-resolution convolutional interaction network to make temporal predictions for cylinder flow data from eddy-viscosity turbulence models in the moderate subcritical regime, but the model is outperformed by basic linear dynamic mode decomposition for a cylinder flow with constant inlet velocity. \citet{fukami2021} combine a convolutional neural network auto-encoder with sparse identification of nonlinear dynamics (i.e. SINDy) to model laminar cylinder flow data generated by direct numerical simulation, as well as a shear flow model. This method is capable of learning the latent dynamics of both cases, however, the resulting models are sensitive to noisy observations and require problem-specific handling of learning parameters. These previous approaches are all limited to predicting numerical simulations of fluid flows, and very few of them include considerations of the impact of noisy or imperfect measurements. Additionally, approximating with standard neural networks means that the learned dynamics can only be evaluated at fixed points used during training. 

Neural operators are a recently developed class of machine learning techniques that are particularly well-poised for applications within fluid mechanics. Neural operators share much of the basic structure of standard neural networks commonly used to approximate functions but are distinct in their ability to approximate operators. Directly approximating operators allow for these methods to learn mappings between infinite-dimensional function spaces (i.e. sets of functions), whereas standard neural networks are typically used to approximate a single function. Learning mappings between infinite-dimensional function spaces enables the approximation of solution operators to systems of partial differential equations (PDEs) such as the constitutive equations underlying physical processes studied in fluid mechanics.  

Introduced by \citet{li2021fno}, Fourier neural operators (FNOs) combine linear transforms in Fourier space with local non-linear activation functions as found in standard neural networks. The linear transforms performed in Fourier space (via the Fast Fourier Transform) serve as global integral operators in real space, which allow FNOs to learn non-local effects for highly non-linear operators.  Because they directly approximate solution operators, FNOs are also discretization invariant meaning that the learned operator can be evaluated on an arbitrary mesh regardless of the discretization of training data. \citet{li2021fno} previously applied FNOs to computational solutions of the Navier-Stokes equation, achieving zero-shot super-resolution and successfully predicting the temporal development of solutions to the vorticity equation. FNOs were also demonstrated on fluid flows in a follow-up work by \citet{li2022fourier}, where they approximated solutions to a wavy pipe flow and a transonic airfoil to demonstrate the approach on general geometries. Other neural operator variations (e.g. DeepONet~\citep{karn2019deeponet}) have been applied to computational flow data as well in different contexts \citep{karn2021boundary,karn2021bubble}. 

Orders of magnitude faster than numerical solvers \citep{kovachki2021neural}, FNOs can be evaluated in just milliseconds which gives them the potential to predict the evolution of fluid flows faster than they occur in real-time. Accurate, full-field predictions could have significant implications for real-world engineering problems such as the mitigation of atmospheric gusts and control of turbulent boundary layers. However previous studies using neural operators on fluid mechanics have focused on problems in computational simulation with perfect information, convenient parameters, and known boundary conditions. To the best of our knowledge, this is the first application of operator learning on experimental measurements of fluid flow. 

Here we explore the potential for FNOs as a real-time-capable machine learning technique for the prediction of fluid flows. We train FNOs on experimental flow data measured via particle image velocimetry (PIV) to forecast the time evolution of cylinder wakes at a range of Reynolds numbers in the subcritical vortex shedding regime. The FNO learns to predict the evolution of the flow over ten time-steps, which are non-dimensionalized so that the resulting prediction horizon is equivalent to roughly 1.7 diameters of translation in the free stream. This relatively short forecast window is chosen to demonstrate FNOs for modeling flow phenomena in the context of predictive control methods with finite planning horizons.  We find that this operator learning approach accurately predicts  instantaneous velocity fields over the full range of Reynolds numbers tested (L2 norm error $< 0.1$) and is robust to experimental noise.



\section{Experimental setup}\label{sec:experimental_setup}

\subsection{Data acquisition}
Our data was collected via experiments performed in a small free-surface water tunnel with a test section of 0.15 m (W) $\times$ 0.15 m (H) and a length of 0.61 m.  We performed tests at flow speeds ranging from $U$ = 0.02 m s$^{-1}$ to $U$ = 0.40 m s$^{-1}$. The corresponding Reynolds numbers range from about $\Rey$ = 240 to $\Rey$ = 3100, where we define Reynolds number as $\Rey  = UD/\nu$ with $D$ being the cylinder diameter and $\nu$ being the kinematic viscosity. The cylinder diameter is held constant at $D$ = 9.53 $\times$ 10$^{-3}$ m and is fully submerged and fixed to the tunnel walls on both sides. The cylinder is made of cast acrylic and is mounted approximately equidistant from the free surface and the tunnel floor to minimize the impact of either boundary on the vortex wake. 

\begin{figure}
  \centerline{\includegraphics{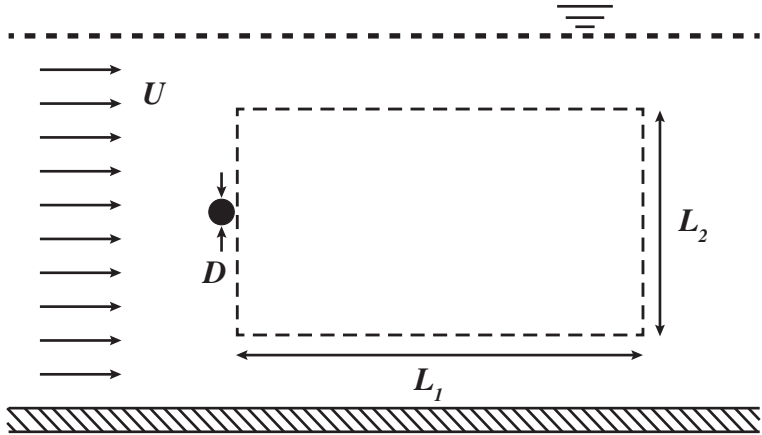}}
  \caption{Schematic of the imaging region of interest, outlined by the black dashed line, relative to the water tunnel and cylinder.}
\label{fig:exp_setup}
\end{figure}

Figure \ref{fig:exp_setup} depicts the region of interest relative to the cylinder and the tunnel boundaries. Here $L_1$ = 0.125 m ($\approx 13D$) and $L_2$ = 0.084 m ($\approx 9D$). This region, located immediately behind the cylinder, is illuminated by a laser sheet at a streamwise cross section near the center of the tunnel. A high-speed camera (IDT XSM-3520 set to 2144 $\times$ 1440 resolution) is used to record the flow in this region. The frame rate of the camera is adjusted based on the mean flow speed to maintain a near-constant non-dimensional time between frames regardless of tunnel speed. The non-dimensional time, otherwise known as the formation time, for a cylinder flow is given by \citet{gharib2004} as: 

\begin{equation}
  t^* = \frac{Ut}{D}
  \label{formation_time}
\end{equation}

Here $t^*$ is the non-dimensional time and $t$ is the dimensional time. 
We used a two-dimensional two-component particle image velocimetry (2D2C-PIV) algorithm to calculate velocity fields by correlating consecutive image pairs. We set an interrogation window size of $32 \times 32$ pixels with 50\% overlap, giving a spatial resolution of $<0.1D$ for our resulting velocity fields. 
Although our measurements only resolve velocity components parallel to the laser sheet, flow in the subcritical vortex shedding regime is known to be three-dimensional \citep{williamson1996review}. We also observed additional three-dimensional effects in the form of an oblique vortex shedding angle, which is known to have implications on the $St-Re$ relationship \citep{hammache1991experimental, prasad1997three}. In addition to the additional challenges associated with increased measurement noise from out-of-plane velocities and a more complicated learning problem (learning the two-dimensional evolution of an inherently three-dimensional flow), this means that matching the non-dimensional time ($t^*$) defined in equation \ref{formation_time} does not guarantee equivalent Strouhal periods. 


\subsection{Fourier neural operators}

Neural operators are distinct from standard neural networks in their unique ability to directly approximate operators, such as mappings between infinite-dimensional function spaces, from data alone. This includes operators that map between sets of functions related by a family of PDEs (e.g. the Navier-Stokes equations), making these methods particularly well-suited for problems in fluid mechanics. . Additionally, the learned solution operator can be evaluated in just a few milliseconds using a standard GPU, making it orders of magnitude faster than the pseudo-spectral method \citet{kovachki2021neural}. 
Introduced by \citet{li2021fno}, Fourier neural operators (FNOs) are a powerful form of neural operator that are guaranteed universal approximation for continuous operators \citep{kovachki2021fno, kovachki2021neural}. They are also mesh-invariant, meaning that they can be trained and evaluated with data of varying resolutions. 

\begin{figure}
  \centerline{\includegraphics[width=0.75\textwidth]{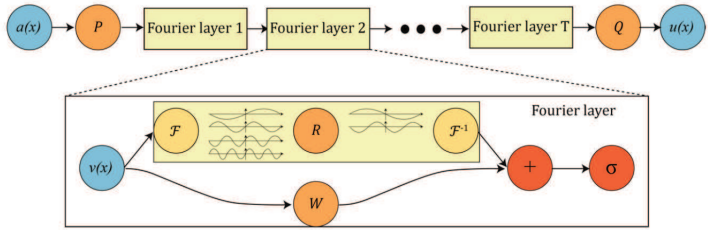}}
  \caption{Diagram showing composition of Fourier layer in FNO. Taken, with permission, from \citep{li2021fno}.}
\label{fig:FNO}
\end{figure}

The architecture of an FNO, shown in figure \ref{fig:FNO}, is made up of unique Fourier layers. These Fourier layers consist of paths: non-linear activation functions as found in classical neural networks, and a linear transform of the input signal performed in Fourier space. The non-linear activation function path serves to approximate local non-linearities, and the transform in Fourier space serves as a global integral operator to account for non-local effects in real space. The paths are then combined and passed forward. The weights in each layer are trained through back-propagation to minimize the selected loss function, in a way that parallels the training of standard neural networks. Details on the mathematical principles underlying FNOs and neural operators more generally can be found in \citet{li2021fno}, \citet{kovachki2021neural}, and \citet{kovachki2021fno}. The FNO hyperparameters used in this work can be found in appendix \ref{appB}.

\section{Results}\label{sec:results}

\subsection{Prediction approach}

In this study, we use FNOs to forecast the time evolution of both $x$ and $y$ components of velocity fields ($u,v$) in the wake of a cylinder at various Reynolds numbers. In this context, FNOs can learn to predict future states of the velocity fields based on current observations. While not necessary for learning with FNOs, we preprocessed the data by subtracting out the mean field. This approach, borrowing from stability theory \citep{landau1987}[pp. 95], essentially decomposed the velocity into steady and unsteady parts: 
\begin{equation}
\boldsymbol{u} (\boldsymbol{x} ,t) = \boldsymbol{u_0}(\boldsymbol{x} ) + \boldsymbol{u'}(\boldsymbol{x} , t)
\label{decomp}
\end{equation}
We then can substitute this into the Navier-Stokes equations. Assuming the steady part alone satisfies the time-independent equations and omitting $\boldsymbol{u'}(\boldsymbol{x}, t)$ terms of order greater than one, we are left with

\begin{equation}
\frac{\partial \boldsymbol{u'}}{\partial t} + (\boldsymbol{u_0} \cdot \nabla) \boldsymbol{u'} + (\boldsymbol{u'} \cdot \nabla) \boldsymbol{u_0} = \frac{\nabla p'}{\rho} + \nu \Delta \boldsymbol{u'}, \ \ \ \  \nabla \cdot \boldsymbol{u'} = 0
\end{equation}
where $p'$ is the unsteady pressure component (i.e. $p(\boldsymbol{x},t) = p_0(\boldsymbol{x}) + p'(\boldsymbol{x},t)$). Therefore, we are left with a set of homogeneous linear differential equations \citep{landau1987}, which are perhaps more readily learned by the FNO. 
\begin{figure}
  \centering
  \includegraphics{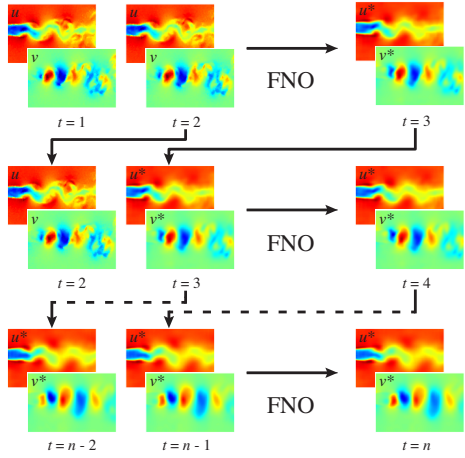}
  \caption{Recursive application of Fourier Neural Operators (FNOs).}
\label{fig:recursive}
\end{figure}

In addition to simplifying the underlying differential equations, subtracting out the time-averaged velocity ensures that unsteady fluctuations are not dominated by the large free-stream velocity bias. This is common to many modal analysis techniques used in studying fluid flows (e.g. proper orthogonal decomposition) \citep{taira2017} and likely has similar benefits in frequency-based learning methods like FNOs. A comparison of performance between FNOs provided the full velocity components ($\boldsymbol{u}(\boldsymbol{x},t)$) and fluctuation velocity components ($\boldsymbol{u'}(\boldsymbol{x},t)$) can be found in the appendix. Since we train the FNOs to predict the fluctuations alone, we can then reconstruct the full flow field by summing the mean and fluctuating velocity components as defined in equation \ref{decomp}.

While FNOs can make predictions from only one time-step as input, we chose to use two time-steps to reduce the effects of experimental errors. From this two time-step input, the model predicts the state of the flow field at the following time-step. During training, we set a desired number of time-steps to forecast. The model is then applied recursively, as shown in figure \ref{fig:recursive}, to reach the desired number of steps. The loss is calculated at each recursive step and summed to find the overall loss in the prediction. The FNO tries to minimize this loss through the back-propagation algorithm, similar to a classic neural network. Because they are applied recursively, the FNO models can also be used to predict time-steps beyond the trained horizon. 

In analyzing predictions, we define the error of the flow field, $\epsilon$, at a given time-step $t$ as calculated as the L2 error norm:
\begin{equation}
    \epsilon(t) = \frac{\| \boldsymbol{q}_t^* - \boldsymbol{q}_t \|_2}{\| \boldsymbol{q}_t \|_2}
    \label{error}
\end{equation}

where $\boldsymbol{q}_t$ is a $N \times 1$ vector containing the two-component velocity field, such that 

\begin{equation}
\setlength{\arraycolsep}{0pt}
\renewcommand{\arraystretch}{1.3}
\boldsymbol{q}_t = \left[
\begin{array}{c}
  u(x_1, y_1, t)  \\
  \displaystyle
  u(x_1, y_2, t)  \\
  ... \\
  u(x_1, y_m, t) \\
  u(x_2, y_1, t) \\
  ... \\
  u(x_n, y_m, t) \\
  v(x_1, y_1, t) \\
  ... \\
  v(x_n, y_m, t) 
\end{array}  \right] 
\label{qvector}
\end{equation}

and $\boldsymbol{q}_t^*$ is the FNO estimate of $\boldsymbol{q}_t$. The quantities $u(x_i,y_j,t)$ and $v(x_i, y_j, t)$ in equation \ref{qvector} are the full, reconstructed velocity components at $(x_i, y_j)$ and time $t$. 

We non-dimensionalized the time-step across Reynolds numbers using the same formulation shown in equation \ref{formation_time}. This time-step was set to be exactly five times the inter-frame time for each Reynolds number ($\Delta t \approx 0.17$). This relatively large time-step was chosen because the measured difference in consecutive velocity fields was on the order which we would expect experimental noise to occur. Using a larger time-step increased the difference while maintaining the same noise level.

\subsection{Prediction accuracy and Reynolds numbers}

We first apply FNOs to cylinder flows at various Reynolds numbers to gauge the performance of the learned solution operators under varying conditions. To this end, we recorded equivalent data sets at $\Rey = \{240, 630, 890, 1260, 1860, 2480, 3060\}$ and trained separate FNOs for each. The velocity fields are predicted in increments of $t^* \approx 0.17$, and the FNO is trained to forecast the evolution of the flow over ten time-steps. 

The Reynolds numbers tested range from the very early transition where three-dimensional effects begin to be observed well into the shear-layer transition regime \citep{williamson1996review}. As the Reynolds number increases, the flow becomes noticeably more irregular which presents challenges for learning the solution operator. As the flow becomes more turbulent the wake becomes three-dimensional and chaotic \citep{karn1992chaos}. Fluctuations in the shape and motion of the shed vortices become irregular, which generally makes forecasting the time evolution of the velocity fields difficult. Intrinsically coupled to the progressively complex physics associated with increasing Reynolds number, growing three-dimensional velocity components move particles in and out of the stream-wise imaging plane which negatively impacts the performance of PIV algorithms. In effect, as the incident velocity increases the learning problem becomes more challenging and the data training becomes noisier. 

\begin{figure}
  \centerline{\includegraphics[width=1\textwidth]{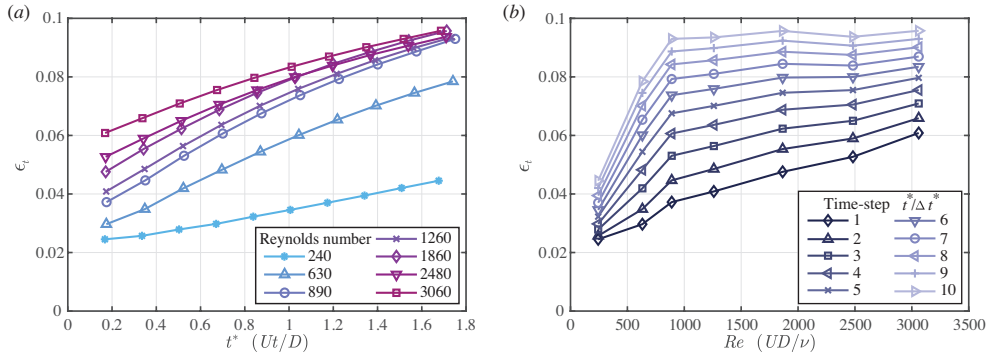}}
  \caption{Relationships between error, non-dimensional time, and Reynolds number. ($a$) The error as a function of prediction time is plotted at various Reynolds numbers. ($b$) The error a function of Reynolds number plotted at various time-steps. }
\label{fig:reynolds}
\end{figure}

From figure \ref{fig:reynolds}a we see that the prediction error does increase with Reynolds number, but remains under 0.10 across the full ten-step prediction horizon, suggesting the FNO is accurately predicting the instantaneous velocity fields even in the more difficult high Reynolds number cases. This means that the total magnitude of the error field after 10 predictions is still less than one-tenth of the magnitude of the total velocity field itself. Figure \ref{fig:reynolds}a also shows that the trajectory of the prediction error does not scale linearly with Reynolds number. Further, we see that the relationship changes depending on the prediction step, as the distribution of the error values at the initial time-step have a very different distribution from those at the final time-step. Plotting the error at each prediction step as a function of Reynolds number (figure \ref{fig:reynolds} b), we see that the error in the first time-step does have a nearly linear relationship with Reynolds number. However, this does not hold and the error begins to grow non-linearly with Reynolds number. The error value of the final time-step ($t^* / \Delta t^* = 10$) increases sharply in the range from $\Rey = 240$ to $890$ but then remains relatively consistent around $\epsilon \approx 9.3 \times 10^-2$. 

\begin{figure}
  \centerline{\includegraphics[width=1\textwidth]{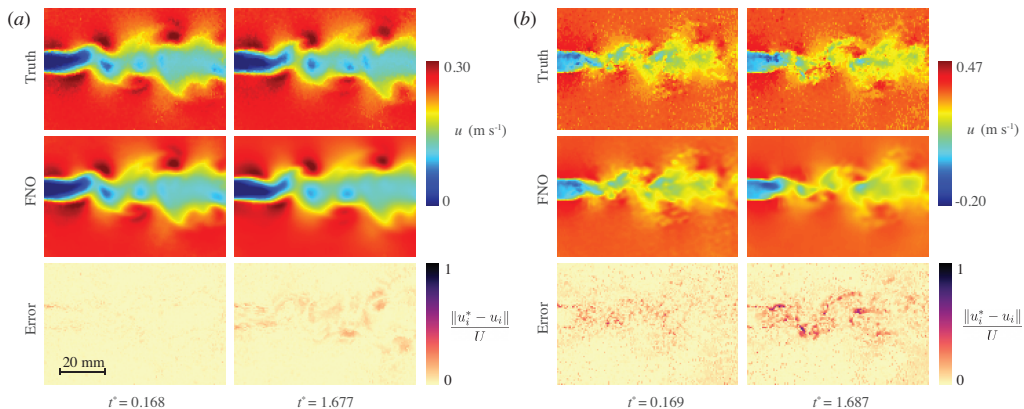}}
  \caption{Instantaneous $x$-velocity component measurement (top row), prediction (middle row), and error (bottom row) at the first (first column) and last (second column) prediction for ($a$) $\Rey = 240$ ($b$) $\Rey = 3060$}
\label{fig:reynolds_fields}
\end{figure}

We can also examine individual true and predicted velocity fields to gain insight into the performance of FNO. Figure \ref{fig:reynolds_fields}ab shows the $x$-component of these values as well as a normalized absolute difference between these two fields at the first and last step for both the $\Rey = 240$ and $\Rey = 3060$. The first step prediction for the $\Rey = 240$ case (left-column, figure \ref{fig:reynolds_fields}a) shows some minor deviations in the outer shear layers as well as small magnitude high-frequency error that seems to concentrate near the edge of the wake. After 10 predictions (right-column, figure \ref{fig:reynolds_fields}a) more differences between the prediction and experiment have developed on the length scale of wake vortices themselves. As would be expected, these regions mostly appear near the edges of wake structures. The measured velocity fields in the $\Rey = 3060$ case (top row, figure \ref{fig:reynolds_fields}b) appear less smooth than the previous case, but the predictions themselves remain smooth. Similar to the low Reynolds number case, the difference between measurement and prediction after the first case primarily consists of high-frequency fluctuation, but we see that these appear to concentrate in the wake region itself. There does appear to be some coherence in the distribution of these regions. We also note that the final prediction for the high Reynolds number case has roughly equivalent high-frequency components in the free-stream portions as the first prediction, but accumulates additional differences in the wake on the scale of the coherent structures similar to the lower Reynolds number case. 

The error regions with length scales on the order of the primary Kármán vortices are likely attributable to the learned FNO model. We see that in both high and low Reynolds number cases, these errors accumulate with increased prediction time. In the $x$-component case shown, they are most notable near the edges of the individual vortices.

The source of the high-frequency differences we observe is less clear. Since we can expect to see increasing three-dimensional velocity fluctuations as the Reynolds number increases in the range shown, it is possible that some of this signal can be attributed to error introduced by out-of-plane motion of seeding particles. While the out-of-plane particle motion is physical, they are not correctly characterized by the two-component PIV algorithm applied here and therefore appear as noise in the resulting measurement. This is most likely the cause of the small deviations in the free-stream regions of the flow, especially for the high Reynolds number case. 

However, it is also possible that some of the high-frequency deviations are directly due to FNO modeling error. We see in both cases that the predicted velocity field becomes smoother between the first and final prediction, especially around the edges of the wake and between low-velocity regions associated with vortex cores. These are regions where we would expect to see the most turbulent fluctuations and small-scale flow features, especially in the high Reynolds number case where the shear layer has transitioned before the vortex formation process begins \citep{williamson1996review}. The increasing smoothness observed in the predicted solutions may therefore indicate that the FNO struggles to model the complicated physics in these turbulent regions, and therefore learns to predict a more average solution absent of small, sharp flow features. We also note that the limited spatial resolution of the PIV measurements ($\approx 0.1D$) is such that not all small structures (e.g. shear layer vortices) will be well resolved in our measurement which may in part make learning these fine-scale dynamics less tractable \citep{williamson1996review,wei1986}.  The more turbulent regions of the flow, however, likely also have more out-of-plane velocity components which then contribute to the aforementioned measurement noise. It is likely a combination of these factors that contributes to the high-frequency component of the error. Despite these minor deviations, we note that FNO does accurately predict the instantaneous shape and motion of the dominant coherent structures in the turbulent wake.

\begin{figure}
  \centerline{\includegraphics[width=0.75\textwidth]{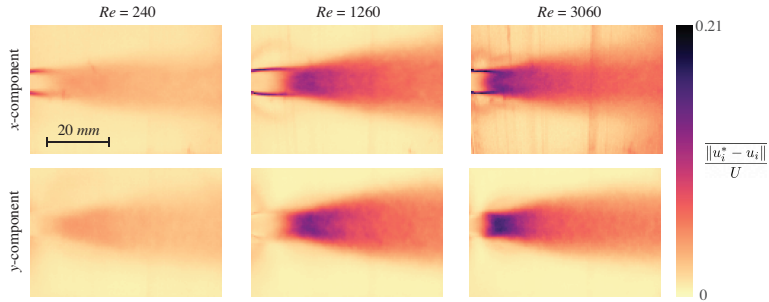}}
  \caption{Mean absolute error field of $x$ and $y$ components normalized by free-stream velocity at different Reynolds numbers.}
\label{fig:normalized_error}
\end{figure}
Figure \ref{fig:normalized_error} shows both components of the absolute time-resolved error averaged over the final time-step of the testing data. In taking the average of the error across many different predictions, we can establish which regions of the wake tend to accumulate the most error. We see that the shear layers from either side of the near-wake are a major source of $x$-component velocity error across all three cases. While there is error in the shear layer in all three cases, we see it increase at the higher Reynolds numbers where the shear layer transition begins.  We note that there are established intermittent shear layer instabilities in cylinder wakes, which helps explain the accumulation of error in this region \citep{prasad1997instability}. Both velocity components for the $\Rey = 1260$ and $3060$ cases also have strong error signals near the end of the recirculation region where the primary Kármán vortices are formed. This central peak error region also expands and diffuses in the downstream wake in both velocity components, which can likely be attributed to a combination of the vortices themselves diffusing and the increased variance in shape/trajectory causing a wider occupancy region for the coherent structures. Both components also have regions of elevated error outside the shear layer near the leading boundary of the frame for the larger of the two Reynolds numbers shown, which can likely be attributed to unknown inlet boundary conditions.

\section{Conclusion}\label{sec:conclusion}

We apply a state-of-the-art operator learning technique (Fourier neural operators) to forecast the time evolution of cylinder wakes at a range of Reynolds numbers in the subcritical regime. We find that the resulting FNOs achieve low errors ($\epsilon < 0.1$) even after ten prediction steps.

We find that error increases with Reynolds number for the first few prediction time-steps, as we would expect for modeling increasingly turbulent flow phenomena. However, we also see that the error in the final prediction step increases only the low Reynolds number cases, reaching a plateau at $\Rey = 890$. This may be due to an accumulating smoothness learned to minimize error from unsteady turbulent dynamics that were not learned. However, it is possible that these fine-scale interactions were not learned due to insufficient spatial resolution. Given sufficient information to better resolve small secondary flow structures known to be present in the wake at higher Reynolds numbers (e.g. shear vortices), the FNO could potentially learn to accurately predict flow across a wider set of length scales. While out-of-plane velocities likely contributed to error in our measurements which also impacted FNO performance,  FNOs are not limited to 2D solutions, and it is possible that they could properly resolve 3D structures if given volumetric three-dimensional flow data.
This work represents a first step towards operator learning in experimental fluid mechanics. While this application required PIV processing that precluded FNO from being applied in real-time here, we have shown that this real-time capable prediction method is robust and accurate when trained with experimental data. While real-time PIV systems do exist \citep{willert2010real,kreizer2010real,varon2019adaptive}, full-field velocity information is not a practical reality for control in most deployed engineered systems. However, flow field reconstruction from sparse measurements has been a focus of many recent works \citep{bright2013,callaham2019robust,loiseau2018sparse,fukami2022machine,fukami2021global,erichson2020shallow,manohar2022sparse}. These reconstruction methods could be used directly in conjunction with the FNO methods applied here to predict the future state of full-field data from a set of distributed sparse sensors. Additionally, recent developments for FNOs have enabled more flexible spatial measurements \citep{li2022fourier}, making the FNO itself a potential end-to-end solution for flow field reconstruction and forecasting the time evolution. 
The development of systems with the ability to predict complicated fluid dynamics faster than their physical realization is an exciting new endeavor in fluid mechanics research, however, operator learning methods also hold significant potential in accelerating CFD solvers as well.



\backsection[Funding]{This work was supported by the National Science Foundation Graduate Research Fellowship under Grant No. DGE‐1745301, Bren endowed chair, Kortschak Scholars, PIMCO Fellows, Amazon AI4Science Fellows, and the Center for Autonomous Systems and Technologies at Caltech.  }

\backsection[Declaration of interests]{The authors report no conflict of interest.}


\backsection[Author ORCIDs]{  P.I. Renn, https://orcid.org/0000-0002-5735-3873; C. Wang, https://orcid.org/0000-0002-8271-5637; S. Lale, https://orcid.org/0000-0002-7191-346X; Z. Li, https://orcid.org/0000-0003-2081-9665; A. Anandkumar, https://orcid.org/0000-0002-6974-6797; M. Gharib, https://orcid.org/0000-0003-0754-4193}


\appendix




\section{FNO Hyperparameters}\label{appB}

\begin{table}[h]
  \begin{center}
\def~{\hphantom{0}}
  \begin{tabular}{lc}
Initial learning rate  & $10^{-3}$ \\
Epochs & 200 \\
Batch size & 16 \\
Train set size & 3488 \\
Test set size & 864 \\
Optimizer & Adam \\
Learning rate scheduler & StepLR \\
Learning rate scheduler step size & 5 \\
Learning rate scheduler decay rate & 0.90 \\
Fourier layer width & 80 \\
Fourier layer modes & 24 \\
  \end{tabular}
  \caption{FNO Hyperparameters}
  \label{tab:kd}
  \end{center}
\end{table}


\bibliographystyle{jfm}
\bibliography{jfm}

\end{document}